\documentclass[12pt,preprint,aps,english,superscriptaddress]{revtex4-1}
\usepackage[T1]{fontenc}
\usepackage[utf8]{inputenc}
\usepackage{amsmath}
\usepackage{amssymb}
\usepackage{cancel}
\usepackage{graphicx}
\usepackage{esint}
\usepackage{color}

\usepackage{setspace}

\usepackage{bbding}

\makeatletter

\@ifundefined{textcolor}{}{%
 \definecolor{BLACK}{gray}{0}
 \definecolor{WHITE}{gray}{1}
 \definecolor{RED}{rgb}{1,0,0}
 \definecolor{GREEN}{rgb}{0,1,0}
 \definecolor{BLUE}{rgb}{0,0,1}
 \definecolor{CYAN}{cmyk}{1,0,0,0}
 \definecolor{MAGENTA}{cmyk}{0,1,0,0}
 \definecolor{YELLOW}{cmyk}{0,0,1,0}
 \definecolor{bblue}{rgb}{0.54,0.81,0.94}
 \definecolor{dred}{rgb}{0.8,0.25,0.33}
}

\usepackage{dcolumn}
\usepackage{bm}
\usepackage{enumerate}

\usepackage[caption=false]{subfig}

\newcommand{\captiontitle}[1]{{\bf #1}}

\@ifundefined{showcaptionsetup}{}{%
\PassOptionsToPackage{caption=false}{subfig}}

\usepackage{babel}
\addto\captionsenglish{}
\renewcommand{\thefigure}{\@arabic\c@figure}

\newcommand{\angstrom}{\mbox{\normalfont\AA}}

\makeatother

\begin{document}
\title{One-dimensional flat bands in twisted bilayer germanium selenide 
}
\date{\today}
\author{D. M. Kennes}
\altaffiliation{These authors contributed equally.}
\affiliation{Institut f\"ur Theorie der Statistischen Physik, RWTH Aachen University and JARA-Fundamentals of Future Information Technology, 52056 Aachen, Germany}
\affiliation{Max Planck Institute for the Structure and Dynamics of Matter, Center for Free Electron Laser Science, 22761 Hamburg, Germany}

\author{L. Xian}
\altaffiliation{These authors contributed equally.}
\affiliation{Max Planck Institute for the Structure and Dynamics of Matter, Center for Free Electron Laser Science, 22761 Hamburg, Germany}

\author{M. Claassen}
\altaffiliation{These authors contributed equally.}
\affiliation{Center for Computational Quantum Physics, Simons Foundation Flatiron Institute, New York, NY 10010 USA}

\author{A. Rubio}
\affiliation{Max Planck Institute for the Structure and Dynamics of Matter, Center for Free Electron Laser Science, 22761 Hamburg, Germany}
\affiliation{Center for Computational Quantum Physics, Simons Foundation Flatiron Institute, New York, NY 10010 USA}
\affiliation{Nano-Bio Spectroscopy Group,  Departamento de Fisica de Materiales, Universidad del Pa\'is Vasco, UPV/EHU- 20018 San Sebasti\'an, Spain}

\maketitle

	\textbf{Experimental advances in the fabrication and characterization of few-layer materials stacked at a relative twist of small angle 
 have recently shown the emergence of flat energy bands. 
As a consequence electron interactions become relevant,
 providing inroads into the physics of strongly correlated two-dimensional systems. Here, we demonstrate 
by combining large scale \textit{ab initio} simulations with numerically exact strong correlation approaches that 
 an effective {\emph{one-dimensional}} system emerges upon stacking two twisted sheets of GeSe, in marked contrast to all Moir\'e systems studied so far. 
This not only allows to study the necessarily collective nature of excitations in one dimension, 
but can also serve as a promising platform to 
scrutinize the crossover from two to one dimension in a controlled setup by varying the twist angle, which provides an intriguing benchmark with respect to theory.  
We thus establish twisted bilayer GeSe as an intriguing inroad into the strongly correlated physics of low-dimensional systems.
}



\section*{Introduction}
Understanding emergent, strongly correlated quantum phenomena 
in complex many-body interacting and low dimensional materials is  one of the main driving forces in modern condensed matter research. 
 Strongly correlated systems are fascinating as they challenge our understanding of quantum mechanics fundamentally, but are also highly relevant to technological advances, such as the quest for room temperature superconductivity, ultra-dense and -fast memory solutions as well as quantum computing platforms \cite{Tokura2017}, to name a few. In this context, the study of low-dimensional systems has revealed a zoo of surprising insights into quantum collective behavior of many-body systems, some of which already find far reaching applications in everyday life, e.g. in computer memory (magnetism) and magnetic resonance imaging techniques (superconductivity).

Recently, twisted bilayer graphene \cite{cao2018a,cao2018b,Yankowitz18,Kerelsky18,Choi19} and other van der Waals materials stacked atop each other at a twist  \cite{Jin18,Xian18,Naik18,Wang19TMD,Scuri19,Kerelsky19,Andersen19} have been proposed as novel  material realizations of two-dimensional correlated physics that afford an unprecedented level of control. Previous studies concentrate on few-layer films featuring a $60^\circ$ or $120^\circ$ rotational symmetry stacked at a twist. By forming a large Moir\'e supercell at small twist angles, a quasi-two-dimensional system  with quenched and tunable kinetic energy scales emerges, thereby drastically enhancing the role of electronic interactions. 

Surprisingly, we report here that if instead we consider layered systems stacked at a small twist angle for which the monolayers have a rectangular lattice with only mirror symmetry, an effectively {\it {one-dimensional}} system with quenched kinetic energy scales (flat bands) emerges. This elevates the concept of Moir\'e  systems to include the broad and exciting realm of one-dimensional quantum systems, which from a theory point of view is ideal to study quantum many-body effects, because powerful theoretical tools (bosonization, tensor network approaches, Bethe Ansatz,... \cite{Delft98,Giamarchi03,Scholwoeck11}) can be employed to obtain a nearly complete picture of its collective nature and effects of strong correlations.  Remarkably, we find that varying the twist angle smoothly interpolates between an effectively one-dimensional and a two-dimensional system at low energies, permitting experimental studies of the dimensional crossover in a clean and controllable manner.

To illustrate this point we perform large-scale \textit{ab initio} based simulations of two sheets of GeSe stacked at a twist, where GeSe belongs to the family of 2D group-IV monochalcogenides \cite{gomes15gs} and has a similar structure as phosphorene [see Fig.~\ref{fig:fig1}(a-b)]. 2D GeSe exhibits high air stability and thin GeSe films down to a monolayer have been studied extensively in experiments for their applications in phototransitors and near-infrared photodetectors \cite{vaughn10gs,xue12gs,mukherjee13gs,zhou17gs,wang17gs,zhao18gs,shengli19gs}. 2D GeSe is also predicted to exhibit giant piezoelectricity \cite{fei15gs,gomes15bgs}, room-temperature ferroelectricity \cite{fei16gs,salvador16gs} and ferroelasticity \cite{wu16gs,qian17gs}, strong visible-light absorbance \cite{shi15gs} and a large bulk photovoltaic effect \cite{neaton17gs}.  This renders GeSe an interesting choice as much prior expertise on the (untwisted) material exist and samples are experimentally available. Furthermore, recently an Eshelby twist has already been realized for GeSe, as well as in the structurally similar system GeS \cite{Liu2019}. 

With extensive \textit{ab initio} calculations, we explicitly demonstrate that a quasi-one-dimensional system emerges for twisted bilayer GeSe at small twist angles, where the degree of ``one-dimensionality'' increases  with decreasing angle. Upon including interactions we show that this system is an effective realization of the so-called ionic Hubbard model. This model has attracted a lot of research attention in the past \cite{Fabrizio99,Wilkens01,Takada01,Kampf03,Manmana04,Tincani09,Loida17}, because it features many interesting prototypical (correlated) phases of matter, including band insulators, Mott insulators, bond density waves and Luttinger liquids, and hosts Ising as well as Kosterlitz-Thouless quantum phase transitions.  As a consequence we find that in twisted GeSe all these different phases of matter can be accessed and their respective phase transitions can be studied in a controllable condensed matter setup.
 We explicitly outline the phase diagram including all the above mentioned phases of matter upon varying the filling (experimentally tunable by gating) as well as the ratio of kinetic and interaction energy scales (tunable by the twist angle) at temperatures accessible within current experimental limitations. Furthermore, twisted bilayer GeSe constitutes a unique system for the controlled study of the crossover between two-dimensional and one-dimensional physics via varying the twist angle using the experimental setup outlined in Ref.~\cite{Ribeiro18}, which can be used to shed light on this interesting regime from an experimental viewpoint in the future. This condensed matter based benchmark system could complement results from more conventional quantum simulation platforms \cite{Scheurer2015,Bloch2012,Tomza2019} in the future in terms of scalability of system size and operation temperature. 
Twisted bilayer GeSe, as we demonstrate, is thus an ideal inroad into the strongly correlated nature of low-dimensional systems. 


\section*{Results}
\subsection*{Density Functional Characterization}
\begin{figure}[t]
\centering
\includegraphics[width=0.9\columnwidth]{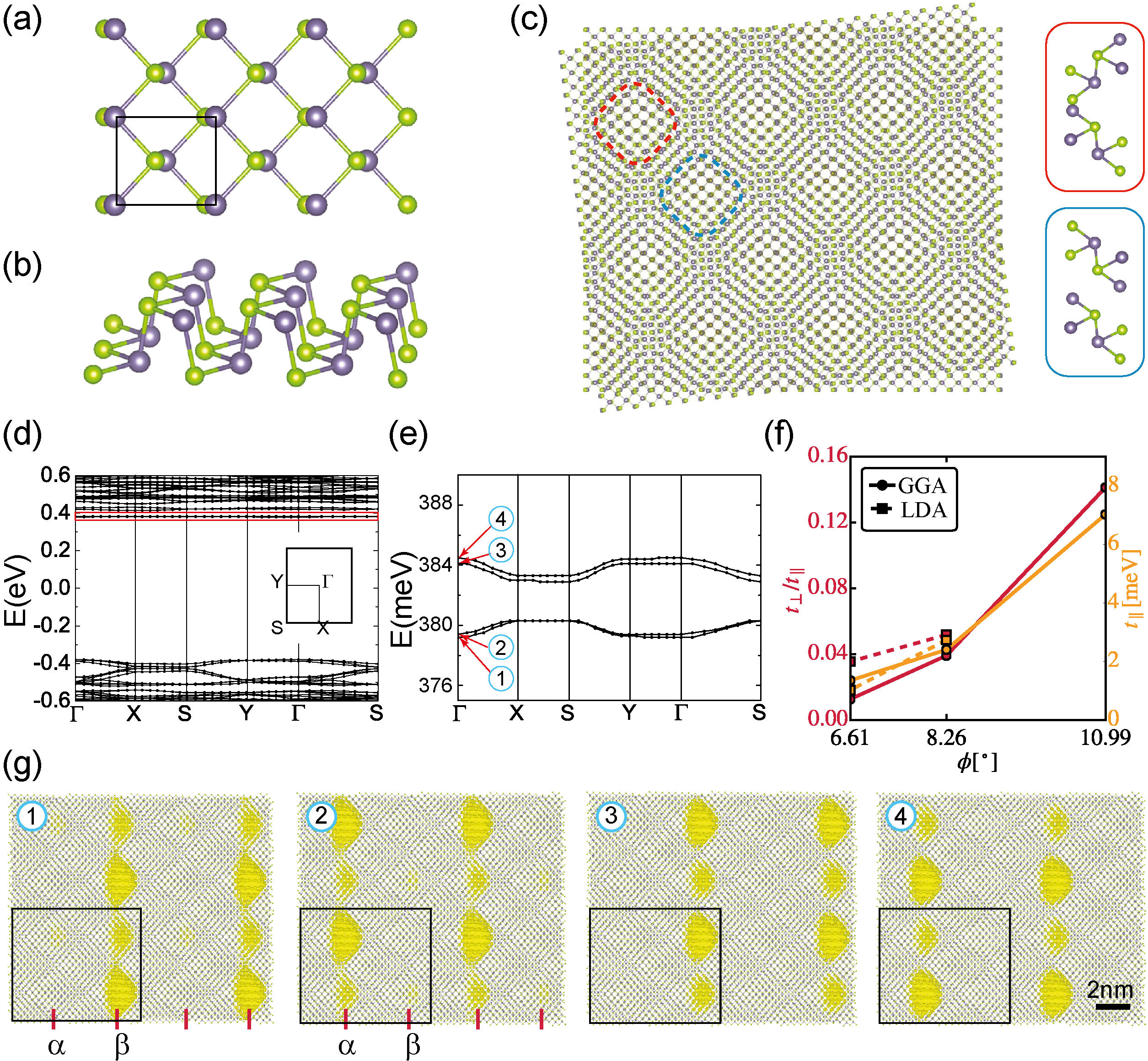}
\caption{\captiontitle{\textit{Ab initio} characterization of twisted bilayer GeSe} (a) and (b) top and side view of monolayer GeSe. Green and blue spheres indicate Se and Ge atoms, respectively. The black box in the top panel denotes the rectangular unit cell of the system. (c) Moir\'e pattern for two sheets of GeSe stacked at a relative twist of 180-6.61 degree denoted by configuration B. The pattern that emerges shows a rectangular shape, with much larger unit cell. We highlight two areas with dashed lines whose staking is given in the right panels.  (d) and (e)  Band structure as obtained from density functional theory using the LDA. Flat bands emerge at the edge of the valence and the conduction bands, where (e) shows a zoom into the red-boxed region highlighted in (d). The flat bands at the conduction band disperse only along one spatial direction, the $\Gamma\to X$ and $S\to Y$ direction. (f) LDA and GGA results for the  ratio between inter-wire ($t_\perp$) and intra-wire ($t_\parallel$) couplings of the emergent one-dimensional chains at low energies as a function of twist angle, highlighting the emergence of quasi-one-dimensional physics at small twists.  (g) Real space illustration of the one-dimensionality of the system showing the charge density of the bands labeled by 1-4 in (e) as accumulated yellow regions (the unit cell hosts a pair of wires with a staggered chemical potential and a wire-wire coupling that vanishes as the angle is decreased). The charge density wires are highlighted with red lines and annotated by $\alpha$ and $\beta$.    }
\label{fig:fig1}
\end{figure}

We start by discussing the \textit{ab initio} band structure results for twisted GeSe. In Fig.~\ref{fig:fig1}  we show the density functional theory (DFT) characterization of two sheets of stacked GeSe at a twist (see methods). The atomic structure of a single sheet of GeSe resides in a rectangular lattice [panels (a) and (b)]. Starting from a perfectly aligned AA-stacking bilayer, different Moir\'e patterns are formed when the top (or the bottom) layer is twisted with angle $\phi$ ranging from $0^\circ$ to $180^\circ$ with respect to the other layer. 

The systems with twist angles $\phi$ and $180^\circ$-$\phi$, which we refer to as configurations A and B, respectively, share supercells of the same size. The supercell for system B is shown in Fig.~\ref{fig:fig1}c, and the corresponding supercell for system A can be found in the method section. We will focus on configuration B in this work and  we refer to the twist angle of $180^\circ$-$\phi$ as $\phi$ for simplicity.
  Similar to the results reported for hexagonal or triangular lattice systems \cite{cao2018a,cao2018b,Yankowitz18,Xian18,Kerelsky18,Naik18,Wu19,Choi19} we find the emergence of flat bands (which as in the case of twisted Boron-Nitride \cite{Xian18} does not rely on tuning to magic angles) at the edges of the conduction and valence bands at small twist angles. However, in marked contrast to these other systems surprisingly some of the low energy bands disperse only along one direction in real space. This is most obvious for bands at the bottom of the conduction bands [see panel (d) and (e)], which are only dispersive along the $\Gamma$-X (or Y-S) direction and dispersionless along the perpendicular $\Gamma$-Y (or X-S) direction.  We carefully checked these results against varying the functionals used in our DFT calculations, which give slightly different relaxed atomic geometry. To this end we compare results obtained within the local density approximation (LDA) to those obtained employing a generalized gradient approximation (GGA) with van der Waals corrections. We find very consistent behavior upon varying the choice of functionals (see methods). Remarkably, we find that the Moir\'e system at small angle shows a quasi-one-dimensional chain-like staggered charge distribution in real space  [see panel (g)] for states in the flat bands, with pairs of wires in the unit cell, each of which displays an alternating sequence of large and small charge puddles. To capture this behavior, we fit the low-energy Moir\'e bands using an anisotropic tight-binding model with a staggered on-site potential (see Methods). Panel (f) summarizes results for such fits obtained within a LDA and GGA. We find (robust to changing the functionals used in DFT) that the ratio between intra- and inter-wire couplings decreases with decreasing twist angle, which tunes the system continuously to the one-dimensional limit.

If we neglect the coupling between the one-dimensional wires at small twist angle, then a simple model that accurately describes the dispersion and charge modulation along the wire is given by a Hamiltonian with nearest-neighbor hopping  $t$ and featuring a staggered on-site potential  $\epsilon_0$
\begin{align}
H_{0,\sigma}&=\sum\limits_i t \; c^\dagger_{i,\sigma}c_{i+1,\sigma}+{\rm H.c.}+\sum\limits_i (-1)^i\epsilon_0 n_{i,\sigma}\label{eq:Hx},
\end{align}
with $n_{i,\sigma}=c_{i,\sigma}^{\dagger}c_{i,\sigma}$ the occupancy at site $i$. The corresponding dispersion has two branches $E_k^{\pm} = \pm\sqrt{4t^2\cos^2(k)+\epsilon_0^2}$.

Starting from the LDA DFT results at a twist angle of $\phi=6.61^\circ$, $\epsilon_0=0.001337$eV can be read off by the gap magnitude at the zone edge, and an optimal fit of the single remaining parameter $t=t_\parallel$ (see methods) is shown in Fig. 2(a). At this angle we find $\epsilon_0/t\approx 1.3$, placing the system in the interesting regime where kinetic energy terms and staggering potential compete in their order of magnitude. More details about the fit as well as the crossover from two dimensions to one can be found in the methods. 


\subsection*{Correlation Effects}

Next, we model electron interaction effects. At this point we have no definite way to pinpoint the range of the interactions and rather adopt the vantage point that screening will promote rather short ranged interactions. To this end we include an on-site repulsion 
with
\begin{equation}
H_U=U\sum_i (n_{i,\uparrow}-1/2)(n_{i,\downarrow}-1/2),
\end{equation}
as the dominant contribution.
The interactions are written in a particle-hole symmetric way for convenience which amounts to an overall shift in chemical potential. This model is known in the literature as the ionic Hubbard model; a paradigmatic model to study the transition from band insulators (BI) to Mott insulators (MI) as the interactions are increased and was investigated extensively at half-filling \cite{Fabrizio99,Wilkens01,Takada01,Kampf03,Manmana04,Tincani09,Loida17}. It is now well understood that this transition occurs via an intermediate bond order wave state (BOW), in which interaction induced spontaneous dimerization leads to alternating strong and weak bonds. The transition from BI to BOW is of the Ising, second order type, while the second transition from the BOW to the MI state is of the Kosterlitz-Thouless (KT) type \cite{Fabrizio99,Wilkens01,Takada01,Kampf03,Manmana04}.  Twisted GeSe thus provides an inroad into this highly intriguing physics and can, depending on the parameters, potentially realize all of these different phases. So far we have used a (zero temperature) \textit{ab initio} analysis of the band structure. For experiments, however, an important question is whether and how the emergent, correlated phases manifest at finite but still low temperature. This can be simulated efficiently for any chemical potential $\mu$ as well as $U$ and $\epsilon_0$ using density matrix renormalization group (DMRG) (see methods) taking the \textit{ab initio} band structure as an input (at higher temperature the band structure itself might be affected but this regime is not the one we focus on in this work).  

\begin{figure}[t]
\centering
\includegraphics[width=\columnwidth]{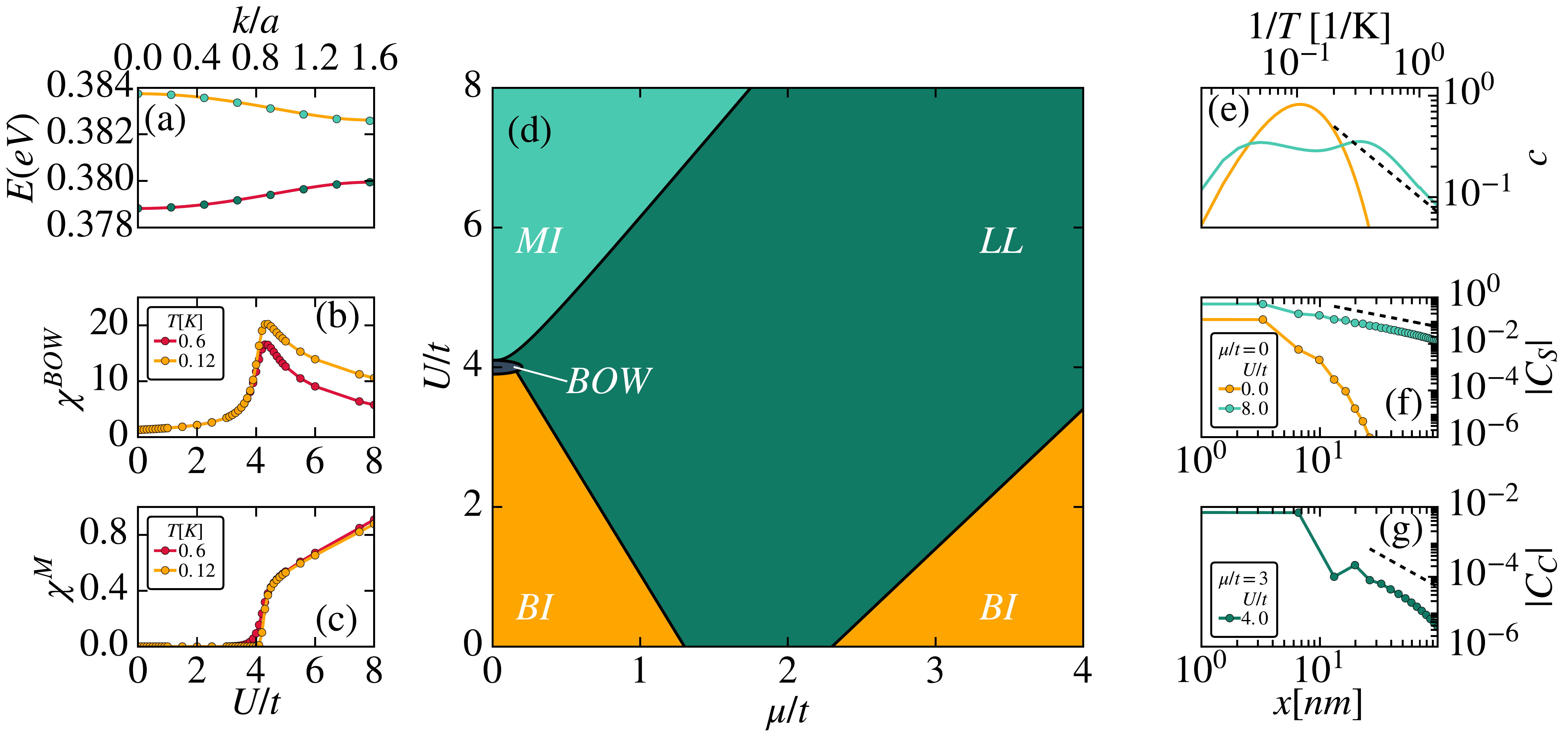}
\caption{\captiontitle{Characterization of many-body electron correlations in twisted bilayer GeSe} (a) Fit (solid lines) to the \textit{ab initio} results shown in Fig.~\ref{fig:fig1}. The fit yields parameters $t=1.03$ meV and $\epsilon_0/t=1.3$ for $\phi=6.61^\circ$. (b), (c) Susceptibilities for bond order as well as magnetization are used to map out the phase boundaries between the Band insulator (BI), the bond ordered wave (BOW) state and the Mott insulating (MI) state at half filling $\mu=0$. The first transition (BI$\to$BOW) is a continuous Ising phase transition, while the second (BOW$\to$MI) is of the Kosterlitz-Thouless type \cite{Fabrizio99,Wilkens01,Takada01,Kampf03,Manmana04}. Upon doping the system away from half filling the system turns to a gapless Luttinger liquid state (at non-zero $U$) characterized by critical power-law correlations in spin and charge degrees of freedom. The full phase diagram at $T=0$ is summarized in (d). (e) Specific heat and (f) spin-spin correlation function at half filling for two values of $U$, placing the system either in the band insulating or Mott insulating state respectively. The specific heat (e) at large inverse temperatures $1/T$ turns from exponential (BI) to linear (MI) which is a hallmark of gapless spin excitations in the MI state. The double maxima structure in $c$ is a hallmark of the lower and upper Hubbard band \cite{Essler09}. We find that at $1K$ the system starts to show clear MI behavior (specific heat $c$ turns linear) for $U/t=8$. Panel (f) shows the spin-spin correlation function. In the BI phase we find exponential suppression, while in the MI state the state shows long range algebraic correlations $C_S\sim x^{-1}$ at $T=1/8K$. Panel (g) shows the charge-charge correlation function obtained for finite doping $\mu/t=3$. The long-ranged power-law decay (dashed line) in the correlation functions falls of as approximately $C_S\sim x^{-1.9}$ which is indicative of a weakly correlated Luttinger liquid (Luttinger parameter $K_C=0.95$) at this $U/t=4$.          }
\label{fig:fig2}
\end{figure}

Much is known about the phases in the ionic Hubbard model  and how to characterize them \cite{Fabrizio99,Wilkens01,Takada01,Kampf03,Manmana04}, particularly at half-filling. We summarize how to distinguish these phases in table \ref{tab:ph}, where we characterize the four different phases band insulator, bond ordered wave, Mott insulator and Luttinger liquid by whether they display a charge gap, spin gap and a staggered bond dimer order. A checkmark signals that the phase displays a non-zero value of the gap or order, while a cross denotes the absence thereof. By calculating the static susceptibility to magnetization $\chi^M$, charging $\chi^C$ and bond ordering $\chi^{BOW}$ upon including a small seed perturbation in magnetic field, onsite potential or bond dimerization, respectively, we determine the spin and charge gaps as well as the bond ordering tendencies (see supplementary note 1). For the smaller angle of $\phi=6.61^\circ$, we show $\chi^M$ and $\chi^{BOW}$ given a small seed $s/t=10^{-2}$ in Fig.~\ref{fig:fig2} (b) and (c). By calculating the static susceptibilities in this fashion and varying $U/t$ as well as $\mu/t$ (corresponding experimentally to controlling the angle as well as back gate) we can map out the phase diagram by using table  \ref{tab:ph}. Panel (d) of Fig.~\ref{fig:fig2} shows the full phase diagram we obtain this way. The BOW state occupies only a tiny fraction of the phase diagram and most likely requires fine tuning to be seen in experiments, especially at finite temperature. 

The different phases of matter manifest prominently in transport experiments with the insulating gap scaling either with $\epsilon_0$ or $U$ in the BI and MI case, respectively, while showing characteristic power-law suppression in temperature in the LL regime. Scanning tunneling  microscopy (STM) will reveal either a charge gap (BI and MI) with different temperature scaling or a power-law suppression of the density of states in the LL case. Both transport and STM have recently been successfully put forward in the twisted van der Waals material's context \cite{cao2018a,cao2018b,Yankowitz18,Kerelsky18,Choi19}. Furthermore, specific heat and spin-spin correlation functions can be monitored to distinguish between these phases. In panel (e) and (f) of Fig.~\ref{fig:fig2} we show the specific heat $c=\partial E/\partial T$ as well as the spin-spin correlation $C_S(x)$ at half filling for two values of $U/t=0$ (BI) and $U/t=8$ (MI). The specific heat in (e) at large inverse temperature $1/T$ is exponentially suppressed in the BI case while for a MI we find a linear behavior which is one of the hallmarks of the emergent gapless spin-excitations. 
We find that at $1K$ the system starts to show clear MI behavior (specific heat $c$ turns linear) for $U/t=8$. Panel (f) depicts the real space spin-spin correlation function. The BI phase is characterized by an exponential suppression of these correlation functions, while one of the hallmarks of the MI state are long range algebraic correlations $C_S\sim x^{-1}$ at $T=1/8K$, at least for small enough distances compared to $1/T$ (after which correlations fall off exponentially). We complement this by studying the charge-charge correlation function $C_C$ obtained for finite doping $\mu/t=3$ shown in panel (g). The long-ranged power-law decay (dashed line) in the correlation functions falls of as approximately $C_S\sim x^{-1.9}$ which indicates a weakly correlated LL state. Importantly, the temperatures for which all of these predictions can be measured are on the Kelvin scale and thus within experimental reach. 

Next, we  highlight the signatures accessible via STM. We compute the density of states $\rho$ at the even lattice sites $i$ by simulating the real time dynamics of $\left\langle c_{i,\uparrow}c^\dagger_{i,\uparrow}(t)\right\rangle$  and taking the Fourier transform. Via the dissipation fluctuation theorem the local density of states can be obtained from this by dividing out the Fermi-distribution $f(-\omega)$ (see Methods for details). The results are summarized in Fig.~\ref{fig:fig3} for temperatures in the Kelvin regime. At small $U$ we find that the single particle gap scales with $\sim\epsilon_0$, while the Mott insulating gap scales as $\sim U$. Overall the behavior of the gap first decreases (with a minimum close to the BOW phase) and then increases as $U$ is increased. The spectral features of the density of states can be used to clearly distinguish experimentally which phases are realized in the system. 

\section*{Discussion}
We have established that twisted bilayer GeSe is an exciting platform to study strongly correlated one-dimensional physics  and the crossover from one to two dimensions in a highly tunable manner. We find that upon marrying \textit{ab initio} materials characterization and strong correlations a one-dimensional ionic Hubbard Model arises, which shows many prototypical features and phases of strongly correlated one-dimensional systems. These can be probed by experiments on twisted bilayer GeSe in accessible temperature regime, albeit on much enlarged Moir\'e length scales. In twisted bilayer GeSe at small twist angles the spin-orbit splitting for the effectively one-dimensional system is negligible.  Future research should address the questions whether in other Moir\'e systems a stronger spin-orbit coupling can be realized. If so this would provide a highly controllable platform to realize Majorana edge state in these effective wires, by coupling the system to a conventional $s$-wave superconducting substrate.

\clearpage

\section*{Methods} 

{\it Details about the DFT Treatment} --- We employed the Vienna Ab initio simulation package (VASP) to perform the ground state DFT calculations \cite{kresse93ab}. The basis was chosen to be plane waves with an energy cutoff of 450 eV and the pseudo potentials are generated using the projector augmented wave method (PAW) \cite{blochl94}. The exchange-correlation functions are treated in the local density approximation (LDA) \cite{perdew81}.  We complement our calculations by also considering the exchange-correlation functionals treated in the generalized gradient approximation (GGA) \cite{pbe} and find results consistent with LDA. A 1x1x1 momentum grid is used for the ground state and relaxation calculations. The experimental lattice constants for bulk GeSe (a=4.38 $\angstrom$, b=3.82 $\angstrom$) are employed for the construction of the supercell of twisted bilayer GeSe. In order to satisfy the commensurate condition, the a lattice constant is slightly expanded by 0.68 $\%$. As periodic boundary condition are applied, a vacuum region larger than 15 $\angstrom$ is added in the z-direction perpendicular to the layers to avoid artificial interaction between the periodic slabs. We relax all the atoms in order to avoid artificial effects as known from unrelaxed structures for other Moire systems \cite{walet2019,lucignano2019,jain2016}. Throughout the relaxation, all the atoms are relaxed until the force on each atom converges to values smaller than 0.01 eV/$\angstrom$. In the GGA calculations, van der Waals corrections are applied using the DFT-D3 method of Grimme \cite{dftd3}. To visualize the charge density distributions of the low-energy states of twisted bilayer GeSe we employ the VESTA code \cite{vesta}. There exist two inequivalent configurations called A and B in the main text, which are illustrated and characterized in Fig.~\ref{fig:S2}. The supercell of twisted bilayer GeSe with twist angles at 10.99$^\circ$, 8.26$^\circ$ and 6.61$^\circ$ contain 872, 1544 and 2408 atoms, respectively.

{ \it Details about the Fitted Band Structure and 1D-2D Crossover} --- We use a simple tight binding model to describe the dispersion at all angles calculated within DFT. We consider a next-nearest-neighbor lattice model on a rectangular grid with a 2 by 2 sites unit cell:
\begin{align}
H_{0}=\sum\limits_i &\phantom{+}t_\parallel \; c^\dagger_{i_x,i_y}c_{i_x+1,i_y}+{\rm H.c.}\notag\\&+t_\perp \; c^\dagger_{i_x,i_y}c_{i_x,i_y+1}+{\rm H.c.}\notag\\&+t_D c^\dagger_{i_x,i_y}c_{i_x+1,i_y+1}+t_D c^\dagger_{i_x,i_y}c_{i_x-1,i_y+1}+{\rm H.c.}\notag\\&+ \epsilon_i n_{i}\label{eq:Hx2D},
\end{align}
with $n_{i,}=c_{i}^{\dagger}c_{i}$ the occupancy at site $i=(i_x,i_y)$.
We fit the dispersion varying the nearest-neighbor hopping amplitudes along the $x$ direction ($t_\perp$), along the $y$ direction ($t_\parallel$), the next-nearest hopping along the diagonal ($t_D$) as well as the onsite potentials $\epsilon_i$. 
We consider a 2 by 2 unit cell so $\epsilon_i$ can take 4 different values $(\epsilon_{0,0},\epsilon_{1,0},\epsilon_{0,1},\epsilon_{1,1})$

Fitting the bands for three different twist angles ${\phi=10.99^\circ}$, ${\phi=8.26^\circ}$  and ${\phi=6.61^\circ}$  yields the values reported in Table~\ref{tab:2new}.
Clearly, as one approaches smaller twist angles the one-dimensional character of the system emerges and the residual chain-chain coupling along the x direction $t_\perp$ and $t_D$ becomes negligible. This is further illustrated in Fig.~\ref{fig:S2Dto1D} where we show the \textit{ab initio} characterization of the dispersion for the same angles as well as the corresponding fits. The bands show more appreciable residual dispersion along the  $X-S$ direction at larger angle, signaling the crossover from 1D to 2D as the angle is increased.  Therefore the effective dimensionality of the system can be tuned by the twist angle and twisted bilayer GeSe provides a tunable platform to study the 2D to 1D crossover.

For the smallest twist angle of $\phi=6.61^\circ$, which we concentrate on in the main text when discussing correlation effects, the dispersion along $x$ is negligible and we can set $t_\perp\approx 0$, $ t_D\approx 0$  as well as label $t=t_\parallel$. Subtracting of the trivial mean potential shift of $\epsilon_{\rm shift}=(\epsilon_{x,0}+\epsilon_{x,1})/2$ and defining $\epsilon_0=(\epsilon_{x,0}-\epsilon_{x,1})/2$ we recover Eq.~\eqref{eq:Hx} of the main text (as well as reinstating the spin degree of freedom).



{ \it Treating Electron Correlations} --- We treat correlations in a numerically exact tensor network based approach formulated in matrix product states \cite{Scholwoeck11}. We exploit the two-site translation invariance of the infinite system and set up the tensor network algorithm directly for the infinite dimensional limit. To treat finite temperature we use the purification scheme described in part 7 of Ref.~\cite{Scholwoeck11} and rewrite the unity operator, corresponding to an infinite temperature density matrix $\rho\sim 1  $ in terms of a wavefunction in combined physical and auxiliary Hilbert space. Subsequently we  ``cool'' the density matrix to temperature $T=1/\beta $, where $\rho\sim e^{-\beta H}$, by applying an imaginary time evolution algorithm. We converge the bond dimension such that numerically exact results are obtained and perform a fourth order Trotter-Suzuki decomposition with small enough steps  in imaginary time $\Delta \beta=0.01$, such that the decomposition does not yield an appreciable approximation. A fourth order decomposition is chosen for numerical convenience allowing for larger time steps then a second order scheme reducing the overall numerical resources needed. In the supplementary note 2 the convergence of all numerical parameters is benchmarked explicitly in the non-interacting limit.

\begin{figure}[t]
\centering
\includegraphics[width=\columnwidth]{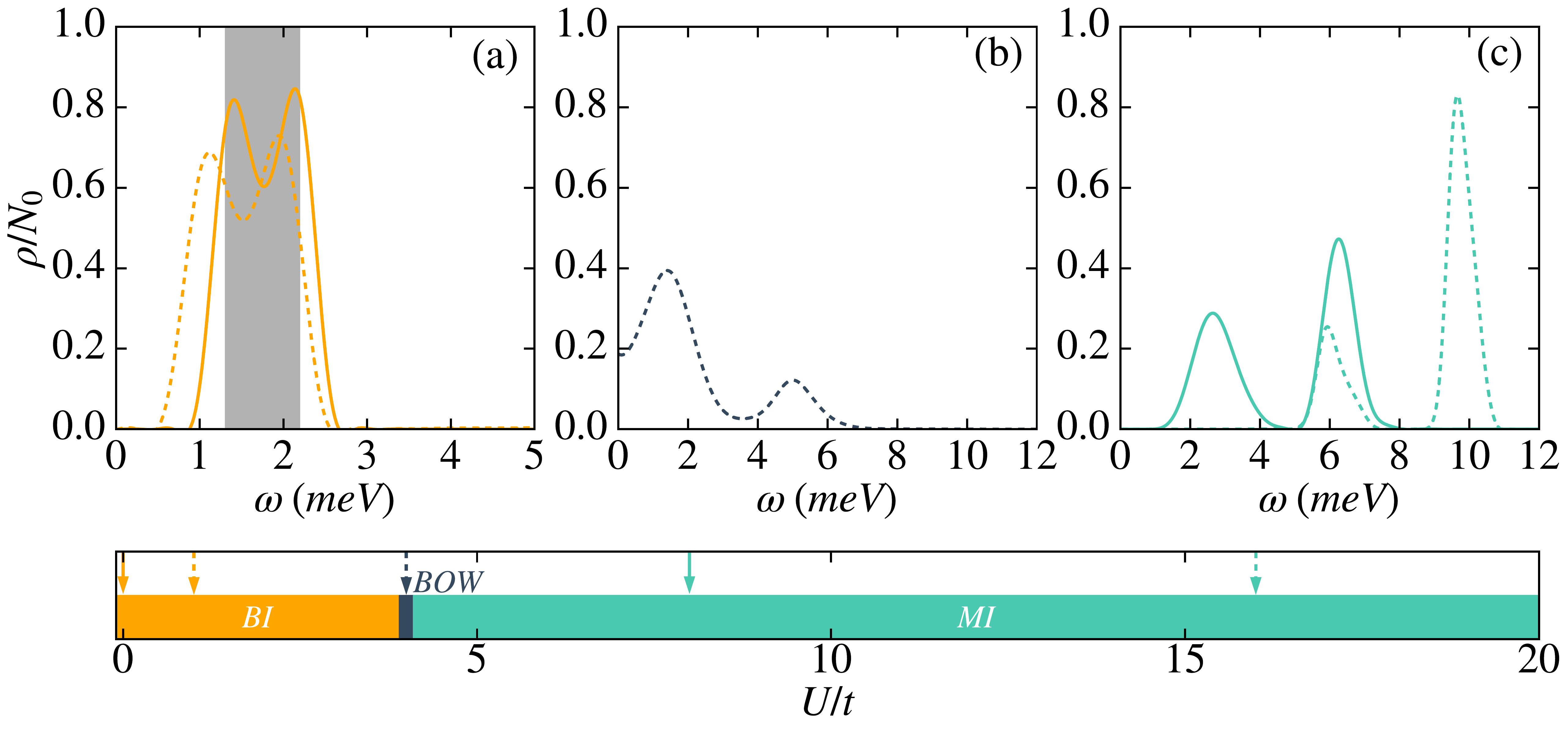}
\caption{\captiontitle{Density of states in twisted bilayer GeSe at $\mu=0$ obtained from DMRG}. The bottom scale shows the different phases found in dependency of $U/t$ at half filling $\mu=0$. Arrows indicated the vales $U/t$ used to calculate the density of states shown in the upper panels ($U/t=0,1$ in (a), $U/t=4$ in (b) and $U/t=8,16$ in (c)), which are grouped corresponding to the phases (BI, BOW or MI in (a),(b) or (c), respectively). In (a) a shaded region indicates the position of the non-interacting band edges, which agrees well with our numerics, where the density of states is found via real-time propagation. Consistent with Fig.~\ref{fig:fig2}, we find a non-monotonic gap size in the density of states as $U/t$ is increases, first decreasing and then increasing. Close to $U=0$ the gap is determined by $\sim\epsilon_0$ while at large $U$ it scales $\sim U$. The temperature in these calculations are $T=1.2 K$ for (a) and (c) as well as $T=2.4 K$ for (b). (Here $N_0$ normalizes the integral over the density of states to one).
}
\label{fig:fig3}
\end{figure}

\begin{figure}[t]
\centering
\includegraphics[width=\columnwidth]{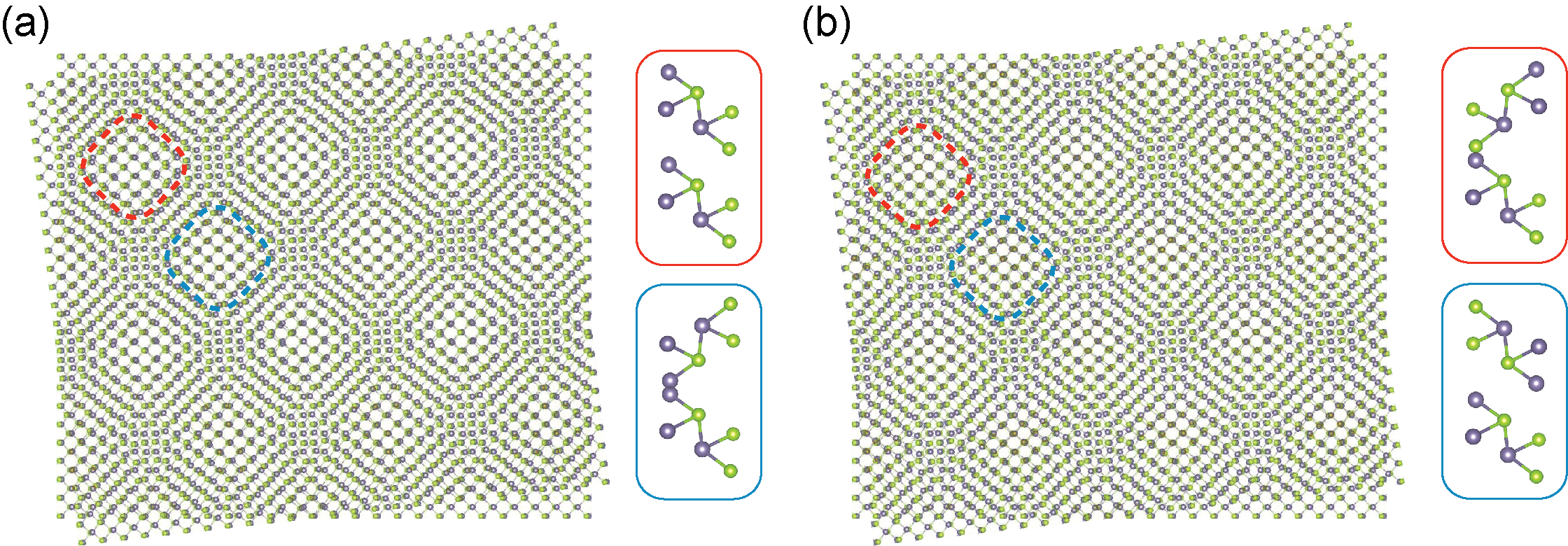}
\caption{The two configurations of twisted bilayer GeSe in real space: (a) configuration A and (b) configuration B. They are related by a $180^\circ$ rotation of the top layer and share the same size of supercell. The insets show the local atomic arrangements in the regions highlighted in red and blue in the main figures.   }
\label{fig:S2}
\end{figure}

\begin{figure}[t]
\centering
\includegraphics[width=\columnwidth]{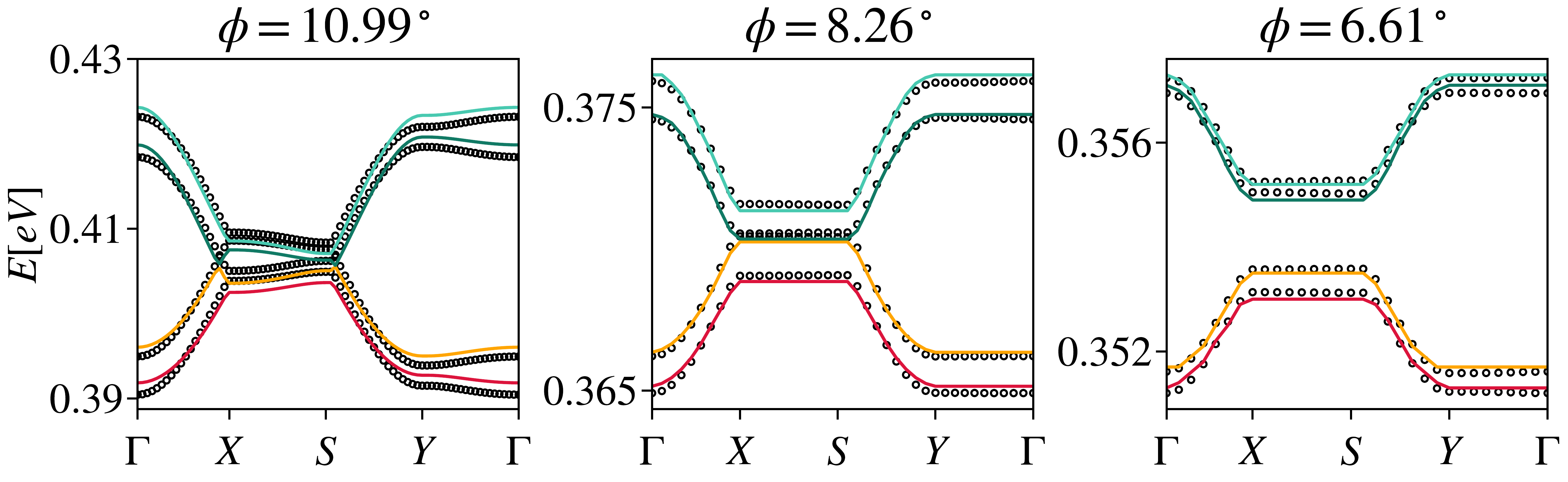}
\caption{ \textit{Ab initio} band structure  obtained with the GGA functionals and including van der Waals corrections (lines) as well as a next-nearest neighbor Hubbard model fit (circles) to the dispersion for different angles. This allows to extract  nearest-neighbor hopping amplitudes along the $x$ direction ($t_\perp$), along the $y$ direction ($t_\parallel$), the next-nearest hopping along the diagonal ($t_D$) as well as the onsite potentials $\epsilon_i$. Clearly the dispersion along the x direction vanishes as we approach smaller angles. The results of the fit are summarized in table \ref{tab:2new}.}
\label{fig:S2Dto1D}
\end{figure}

\begin{figure}[t]
\centering
\includegraphics[width=\columnwidth]{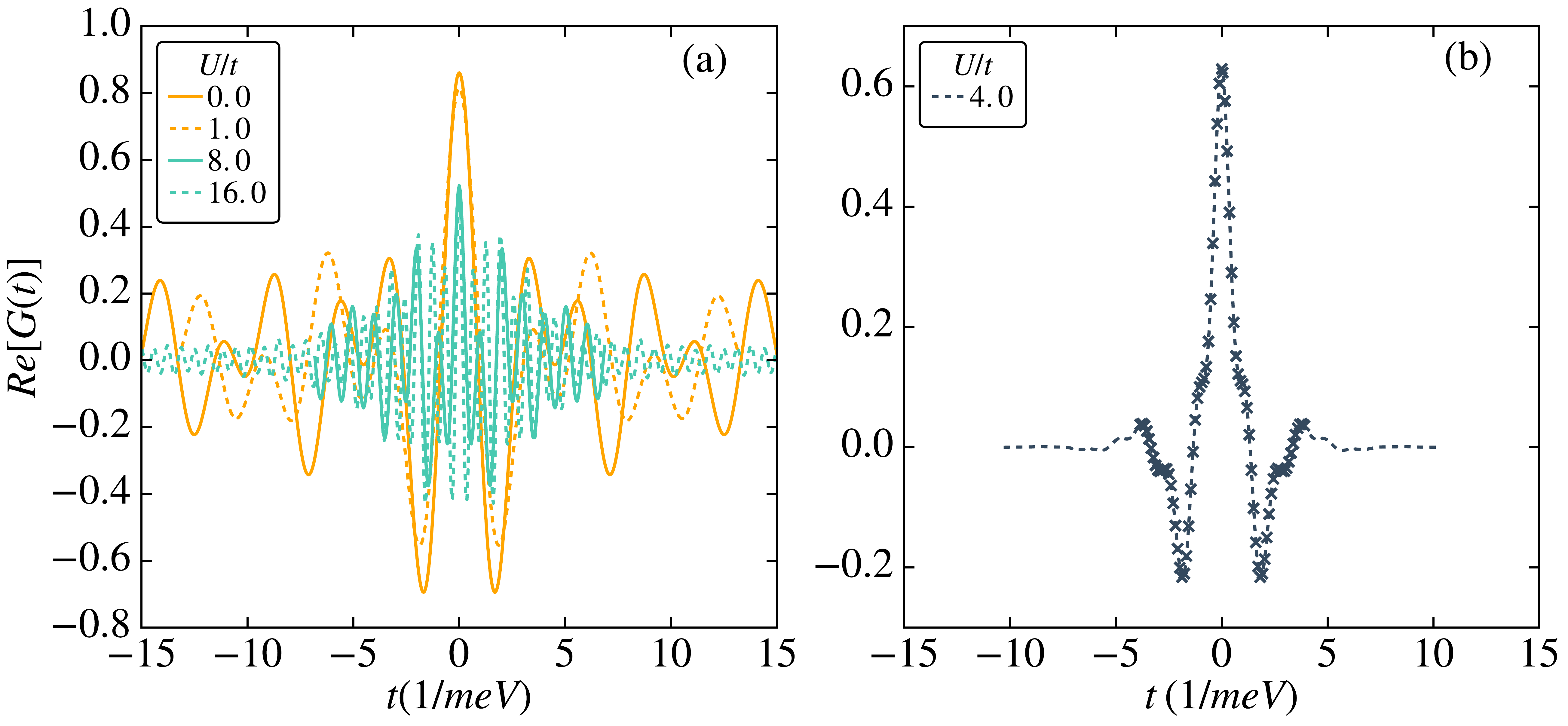}
\caption{Real time simulation of $G(t)$ (see Eq.~\eqref{eq:defDOS}) of the data shown in Fig.~\ref{fig:fig3}. In the case of $U=4$ (shown in (b)) we extend the reached time scales by using linear prediction. Symbols are calculated data points, the line is the data obtained using linear prediction.}
\label{fig:S3}
\end{figure}

{\it Calculating the Density of States} --- To calculate the density of states we use a simulation in real time (and at finite temperature) to obtain the \begin{equation}
 G(t)=\left\langle c_{i,\uparrow}c^\dagger_{i,\uparrow}(t)\right\rangle   \label{eq:defDOS}
\end{equation}. For this we use the ideas put forward in Ref.~\cite{KENNES201637}. This is essential to reach long enough times, such that a meaningful Fourier transform can be taken with a Hanning type window function, compare Fig.~\ref{fig:S3} (a). The maximum time reached by the simulation thus limits the frequency resolution and introduces natural broadening in the Fourier transform. This procedure is employed for the Data shown in Fig.~\ref{fig:fig3} (a) and (c) where the $U/t$ is either large or small both cases in which the entanglement growth is quite moderate. For the data shown Fig.~\ref{fig:fig3} (b) which is $U/t=4$ the entanglement growth is much more severe and even after employing the ideas of Ref.~\cite{KENNES201637}, the time scales are limited. To this end we utilize a linear prediction algorithm to extend the time scales, see Fig.~\ref{fig:S3} (b).

\emph{Data availability:}
All data generated and analysed during this study are available from the corresponding author upon reasonable request.

\emph{Code availability:}
All the custom codes used in this study are available from the corresponding author upon reasonable request.

\emph{Acknowledgments.}
This work was supported by the European Research Council (ERC-2015-AdG694097) and Grupos Consolidados (IT578-13). The Flatiron Institute is a division of the Simons Foundation. LX acknowledges the European Unions Horizon 2020 research and innovation programme under the Marie Sklodowska-Curie grant agreement No. 709382 (MODHET). MC is supported by the Flatiron Institute, a division of the Simons Foundation. DMK  acknowledges  funding by the Deutsche Forschungsgemeinschaft (DFG, German Research Foundation) under Germany's Excellence Strategy  - Cluster  of  Excellence  Matter  and  Light  for Quantum Computing (ML4Q) EXC 2004/1 - 390534769. Gef\"ordert  durch  die  Deutsche  Forschungsgemeinschaft(DFG)  im  Rahmen  der  Exzellenzstrategie  des  Bundes und der L\"ander - Exzellenzcluster Materie und Licht f\"ur Quanteninformation (ML4Q) EXC 2004/1 - 390534769. DMK acknowledges  funding by the Deutsche Forschungsgemeinschaft (DFG, German Research Foundation) under RTG 1995. We acknowledge support by the Max Planck Institute - New York City  Center for Non-Equilibrium Quantum Phenomena.

DMRG calculations were performed with computing resources granted by RWTH Aachen University under projects prep0010. We acknowledge computing resources from Columbia University's Shared Research Computing Facility project, which is supported by NIH Research Facility Improvement Grant 1G20RR030893-01, and associated funds from the New York State Empire State Development, Division of Science Technology and Innovation (NYSTAR) Contract C090171, both awarded April 15, 2010.

In the final step of writing this manuscript arXiv:1905.02206 appeared, which supports the message of this paper.

\hspace{0.5in}

\bibliographystyle{naturemag}
\bibliography{TwistedGESE}

\newpage

\emph{Author contribution:}
 LX provided DFT results, DMK performed the DMRG calculations. DMK, LX, MC and AR interpreted the results and participated in writing the manuscript.

\emph{Competing interests:}
The authors declare no competing interests.

\paragraph{Corresponding author:}
Dante Kennes (Dante.Kennes@rwth-aachen.de) and Angel Rubio (angel.rubio@mpsd.mpg.de)

\newpage

\clearpage

\clearpage

\clearpage

\begin{table}[t!]
\begin{tabular}{ c | c|  c|  c| c}
          & Band Insulator  & Bond Ordered Wave & Mott Insulator & Luttinger Liquid\\
  Charge Gap& {\color{black}\CheckmarkBold} &{\color{black}\CheckmarkBold} &  {\color{black}\CheckmarkBold} & {\color{black}\XSolidBrush}\\
Spin Gap  & {\color{black}\CheckmarkBold} &{\color{black}\CheckmarkBold} & {\color{black}\XSolidBrush} &{\color{black}\XSolidBrush}\\
Bond Dimer& {\color{black}\XSolidBrush}   &{\color{black}\CheckmarkBold} & {\color{black}\XSolidBrush} &{\color{black}\XSolidBrush}\\
\end{tabular}
\caption{Theoretical characterization of the different phases of matter that can be realized in twisted GeSe \cite{Fabrizio99,Wilkens01,Takada01,Kampf03,Manmana04}. The four different phases band insulator, bond ordered wave, Mott insulator and Luttinger liquid are distinguished by whether they display a charge gap, spin gap and a staggered bond dimer order. A checkmark signals that the phase displays a non-zero value of the gap or order, while a cross denotes the absence thereof.}
\label{tab:ph}
\end{table}

\begin{table}[t!]
\setlength{\tabcolsep}{10pt} 
\renewcommand{\arraystretch}{1.5} 
\begin{tabular}{ c|  c  c c c c c c }
$\phi$ &$t_\parallel$&$t_\perp$&$t_D$&$\epsilon_{0,0}$[eV]&$\epsilon_{0,1}$[eV]&$\epsilon_{1,0}$[eV]&$\epsilon_{1,1}$[eV]\\
\cline{1-8}
10.99&7.024&0.992&0.039&408.359&406.214&407.522&404.923\\
8.26&2.403&0.094&0.037&369.076&370.521&370.474&371.556\\
6.61&1.349&0.017&0.036&353.57&355.048&355.249&353.135
 \end{tabular}
\caption{{ Fitted values for the tight binding model of Eq.~\eqref{eq:Hx2D} for three different angles as shown in Fig.~\ref{fig:S2Dto1D}.} Clearly the system becomes more one-dimensional as the angle becomes smaller.}

\label{tab:2new}
\end{table}

\end{document}


\title{Supplementary Information -- One-dimensional flat bands in twisted bilayer germanium selenide}
\date{\today}
\author{D. M. Kennes}
\affiliation{Institut f\"ur Theorie der Statistischen Physik, RWTH Aachen University and JARA-Fundamentals of Future Information Technology, 52056 Aachen, Germany}
\affiliation{Max Planck Institute for the Structure and Dynamics of Matter, Center for Free Electron Laser Science, 22761 Hamburg, Germany}
\author{L. Xian}
\affiliation{Max Planck Institute for the Structure and Dynamics of Matter, Center for Free Electron Laser Science, 22761 Hamburg, Germany}
\author{M. Claassen}
\affiliation{Center for Computational Quantum Physics, Simons Foundation Flatiron Institute, New York, NY 10010 USA}
\author{A. Rubio}
\affiliation{Max Planck Institute for the Structure and Dynamics of Matter, Center for Free Electron Laser Science, 22761 Hamburg, Germany}
\affiliation{Center for Computational Quantum Physics, Simons Foundation Flatiron Institute, New York, NY 10010 USA}
\affiliation{Nano-Bio Spectroscopy Group,  Departamento de Fisica de Materiales, Universidad del Pa\'is Vasco, UPV/EHU- 20018 San Sebasti\'an, Spain}

\maketitle


\section*{Supplementary Note 1: Characterization of Phases}
In the main text we use small seed fields to efficiently characterize susceptibilities towards the different ordering tendencies of the Ionic Hubbard model. Here we present the details of the calculation for completeness. We define the susceptibility
\begin{equation}
\chi^X=O/s
\end{equation}
as the ratio between an appropriately chosen observable (measuring the symmetry breaking accompanied by the phase) and the strength $s$ of a symmetry breaking seed field $\Delta H^X$ added to the Hamiltonian. For the magnetization and charge susceptibilities $X=M$ and $X=C$ we chose $O$ as the magnetization  $M=\sum_{i,\sigma}(-1)^\sigma n_{i,\sigma}  /N$ or charge $C=\sum_{i,\sigma} n_{i,\sigma}  /N$. The seeds added to the Hamiltonian are  $\Delta H^M=s\sum_{i,\sigma}(-1)^\sigma n_{i,\sigma} $ and $\Delta H^C=s\sum_{i,\sigma} n_{i,\sigma} $. For the susceptibility to BOW ordering $X=\textit{BOW}$ we chose $O$ as the dimerization in the hopping  $B=\sum_{i,\sigma}(-1)^i c^\dagger_{i,\sigma}c_{i+1,\sigma}  /N$ and the seed as $\Delta H^\textit{BOW}=s\sum_{i,\sigma}(-1)^i c^\dagger_{i,\sigma}c_{i+1,\sigma}  $.
\section*{Supplementary Note 2: Benchmarking the DMRG with exact solutions}
 In supplementary figure \ref{fig:S1} we benchmark our thermodynamic limit finite temperature DMRG results against exact results obtained in the non-interacting limit $U=0$ of equation (1) in the main text. We calculate the specific heat (as in the main text), the average occupancy $\bar{n}=\lim_{N\to \infty}\sum_i n_i/N$  as well as the difference in occupancy between  even and odd lattice sites $\Delta n=\lim_{N\to \infty}\sum_i (-1)^i n_i/N$ . We show that we can converge the numerical parameters to obtain results which are numerically exact.
\begin{figure}[t]
\centering
\includegraphics[width=\columnwidth]{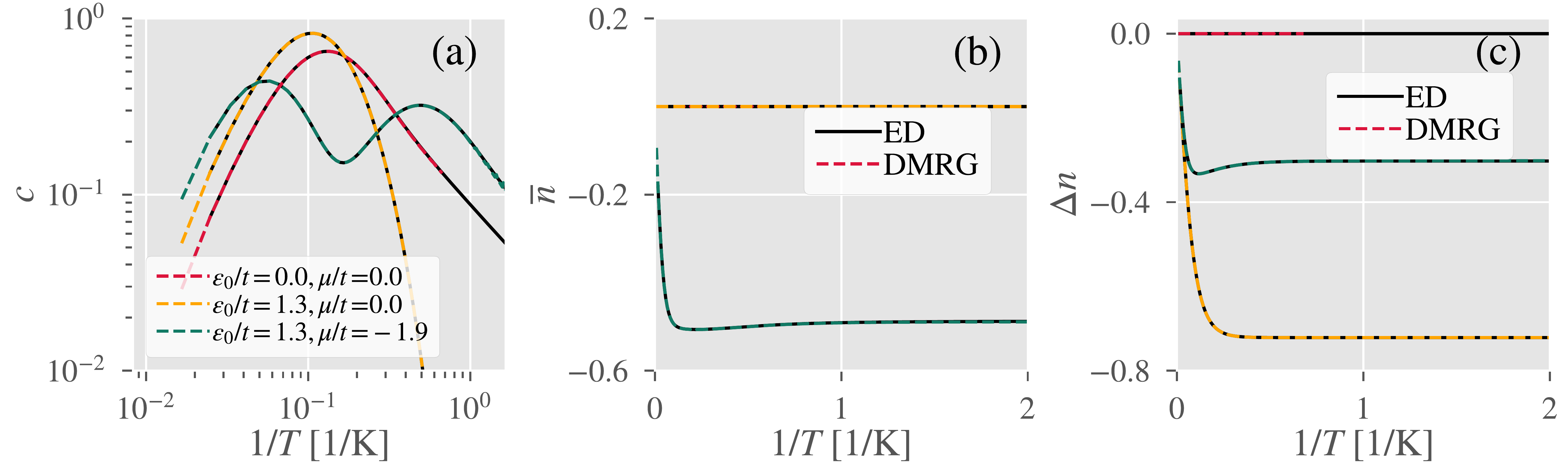}
\caption{{Benchmarking the DMRG against exact solution (ED) at $U=0$} (a) specific heat (b) average occupancy (c) difference in occupancy between  even and odd lattice sites. We show that numerical convergence to the obtain results which are numerically exact can be achieved.   }
\label{fig:S1}
\end{figure}

\bibliographystyle{naturemag}
\bibliography{TwistedGESE}